\documentclass[superscriptaddress,pre,floatfix,twocolumn,amsmath,amssymb,aps,nofootin
,showkeys,showpacs]{revtex4-1}

\usepackage{graphicx}
\usepackage{amsmath}
\usepackage{amssymb}
\usepackage{dcolumn}
\usepackage{color}
\usepackage{multirow}
\usepackage{bm}
\usepackage{subfigure}
\usepackage{tabularx}
\usepackage{booktabs}
\usepackage{appendix}
\usepackage{tikz}

\begin{document}
\title{Revisiting the Fermion Sign Problem from the Structure of Lee-Yang Zeros.\\ I. The Form of Partition Function for Indistinguishable Particles and Its Zeros at 0~K}
\author{Ran-Chen He}
\email{These authors contributed equally to this work.}
\affiliation{Interdisciplinary Institute of Light-Element Quantum Materials, Research Center for Light-Element Advanced Materials, and Collaborative Innovation Center of Quantum Matter, Peking University, Beijing 100871, P. R. China}
\affiliation{State Key Laboratory for Artificial Microstructure and Mesoscopic Physics, Frontier Science Center for Nano-optoelectronics and School of Physics, Peking University, Beijing 100871, P. R. China}
\author{Jia-Xi Zeng}
\email{These authors contributed equally to this work.}
\affiliation{State Key Laboratory for Artificial Microstructure and Mesoscopic Physics, Frontier Science Center for Nano-optoelectronics and School of Physics, Peking University, Beijing 100871, P. R. China}
\author{Shu Yang}
\affiliation{State Key Laboratory for Artificial Microstructure and Mesoscopic Physics, Frontier Science Center for Nano-optoelectronics and School of Physics, Peking University, Beijing 100871, P. R. China}
\author{Cong Wang}
\email{frankcongwang@pku.edu.cn}
\affiliation{State Key Laboratory for Artificial Microstructure and Mesoscopic Physics, Frontier Science Center for Nano-optoelectronics and School of Physics, Peking University, Beijing 100871, P. R. China}
\author{Qi-Jun Ye}
\email{qjye@pku.edu.cn}
\affiliation{Interdisciplinary Institute of Light-Element Quantum Materials, Research Center for Light-Element Advanced Materials, and Collaborative Innovation Center of Quantum Matter, Peking University, Beijing 100871, P. R. China}
\affiliation{State Key Laboratory for Artificial Microstructure and Mesoscopic Physics, Frontier Science Center for Nano-optoelectronics and School of Physics, Peking University, Beijing 100871, P. R. China}
\author{Xin-Zheng Li}
\email{xzli@pku.edu.cn}
\affiliation{Interdisciplinary Institute of Light-Element Quantum Materials, Research Center for Light-Element Advanced Materials, and Collaborative Innovation Center of Quantum Matter, Peking University, Beijing 100871, P. R. China}
\affiliation{State Key Laboratory for Artificial Microstructure and Mesoscopic Physics, Frontier Science Center for Nano-optoelectronics and School of Physics, Peking University, Beijing 100871, P. R. China}
\affiliation{Peking University Yangtze Delta Institute of Optoelectronics, Nantong, Jiangsu 226010, P. R. China}
\date{\today}

\begin{abstract}
    To simulate indistinguishable particles, recent studies of path-integral molecular dynamics formulated their partition function $Z$ as a recurrence relation involving a variable $\xi$, with $\xi=1$(-1) for bosons (fermions).
    Inspired by Lee-Yang phase transition theory, we extend $\xi$ into the complex plane and reformulate $Z$ as a polynomial in $\xi$.
    By analyzing the distribution of the partition function zeros, we gain insights into the analytical properties of indistinguishable particles, particularly regarding the fermion sign problem (FSP).
    We found that at 0~K, the partition function zeros for $N$ particles are located at $\xi=-1$, $-1/2$, $-1/3$, $\cdots$, $-1/(N-1)$.
    This distribution disrupts the analytic continuation of thermodynamic quantities, expressed as functions of $\xi$ and typically performed along $\xi=1\to-1$, whenever the paths intersect these zeros.
    Moreover, we highlight the zero at $\xi = -1$, which induces an extra term in the free energy of the fermionic systems compared to ones at other $\xi=e^{i\theta}$ values.
    If a path connects this zero to a bosonic system with identical potential energies, it brings a transition resembling a phase transition. 
    These findings provide a fresh perspective on the successes and challenges of emerging FSP studies based on analytic continuation techniques.
\end{abstract}


\maketitle
\section{Introduction}
Describing properties of a quantum many-body system is a task of fundamental importance in modern physics, central to disciplines from condensed matter physics~\cite{Bruus_2004,Coleman_2015,RevModPhys.73.33} to quantum computing~\cite{Lloyd_1996,Huggins_2022}, from warm 
dense matter physics~\cite{PhysRevLett.125.078501,DORNHEIM20181,Ernstorfer_2009} to ultracold atomic physics~\cite{RevModPhys.80.885,Douglas_2015}.
Over the past decades, several methods have been developed to tackle this problem.
Among them, density-functional theory~(DFT) provides the most practical solution for a wide range of realistic systems~\cite{Hohenberg_1964,Kohn_1965},
substantially impacting modern physics, chemistry, biology, and other branches of natural sciences.
However, a roadmap toward a systematic improvement of its accuracy is still absent~\cite{Jones_2015}, seriously restricting its application to some ``challenging'' systems, e.g. the strongly correlated ones.
For such systems, the wavefunction methods are often resorted to, 
e.g. exact diagonalization (ED)~\cite{Dagotto_1994}, quantum Monte-Carlo~(QMC)~\cite{RevModPhys.73.33}, density matrix renormalization 
group~(DMRG)~\cite{White_1992,Schollwock_2005}, tensor network~\cite{Orus_2014}, dynamical mean-field theory~(DMFT)~\cite{Georges_1996}, and 
variational quantum eigensolver~(VQE)~\cite{Kandala_2017}, each having its own strengths and limitations.
The exponential growth of the Hilbert space with system size and the intrinsic fermion sign problem (FSP) pose the 
underlying grand challenges.
Parallel to these DFT and the wavefunction based methods, the path-integral representation of quantum mechanics presents an alternative for descriptions 
of a quantum many-body system~\cite{Feynman_1965}.
Its unique advantages for descriptions of the kinematics of a many-body system at the quantum mechanical level make it highly feasible for numerical samplings.
Starting from the early 1980s~\cite{Chandler_1981,parrinello1984}, when combined with the molecular dynamics (MD) and Monte-Carlo (MC) methods, the path-integral based methods have archieved great successes in the community of 
molecular simulations~\cite{Tuckerman_1997,Marx_1999,Habershon_2009,Li_2011,RevModPhys.84.1607,Chen_2013,Richardson_2016,Zhu_2022}. 
Here, the central idea is that the partition function of an $N$-particle quantum system is isomorphic to that of fictitious classical polymers at finite 
temperatures ($T$s)~\cite{Chandler_1981}.
As such, thermodynamic properties of a quantum many-body system can be simulated by path-integral molecular dynamics (PIMD) or path-integral Monte-Carlo (PIMC)~\cite{Tuckerman_1997,Marx_1999,RevModPhys.84.1607}, with the \textit{curse of dimensionality} circumvented~\cite{Li_2014}.
One restriction of such treatments, however, resides on the tag of ``distinguishable'' particles.
Therefore, the ``curse of dimensionality'' is temporarily but not really circumvented~\cite{cod_and_fsp}, and the FSP persists.
A prerequisite for solving or at least partially solving the FSP problem is that in the PIMD/PIMC simulations the particles must be ``indistinguishable''.
Along this route, quite a few effective path-integral methods have been developed in recent years, including the 
worm algorithm in PIMC~\cite{Ceperley1995,Prokofev1998,Prokofev2004,Boninsegni2006},
the recurrence relation in PIMD of bosons~\cite{Hirshberg2019,Myung2022,Feldman2023}, and the re-weighting method at moderate $T$s for fermions~\cite{Hirshberg2020}. 
Satisfactory performances, however, are restricted to the bosonic systems~\cite{RevModPhys.73.33,Huggins_2022}.
For the fermionic systems, using a real parameter $\xi$ to analytically continuate the results obtained from a ``designed'' bosonic system to the targeted 
fermonic one is a popular choice~\cite{RN511,Xiong2023,Dornheim2023,RN534,dornheim2024a,RN546}.
At high $T$s, the numerical performance of such analytic continuations are satisfactory.
However, at low $T$s, the situation becomes much worse.
Besides this, a fundamental question concerning under which circumstances can this approach work, is also unclear.
This analytic continuation is a classic technique in quantum and statistical mechanics, with the Kramers-Kr\"onig relation~\cite{Kronig_1926,Kramers_1927} and the 
Lee-Yang (LY) phase-transition theory~\cite{Yang_1952,Lee_1952} being two prototypical examples.
Here, the analytic properties of the function to be analyzed on the complex plane of its variable, i.e. the dielectric function in the Kramers-Kr\"onig 
relation on the complex plane of the frequency and the partition function in the LY phase transition theory on the complex plane of the thermodynamic state functions, 
are crucial.
Taking the LY phase transition theory as an example, the LY zeros which approach the real axis of the thermodynamic state function at the thermodynamic limit define 
phase transitions, where the analytic properties of the physical quantities as a function of the thermodynamic 
state function are broken~\cite{Yang_1952,Lee_1952}.
It is only when the LY zeros move away from the real axis of the thermodynamic state functions, can the physical quantities as a function of the thermodynamic state function
become analytic again, e.g. as Lee and Yang have demonstrated for the systems when there is no phase transition in the 1950s~\cite{Lee_1952} and as we have shown for 
the supercritical matter last year~\cite{Ouyang_2023}.
However, to the best of our knowledge, the distribution of the partition function zeros has never been discussed in the existing theoretical/numerical 
studies of the FSP in PIMD/PIMC~\cite{Ceperley1995,Prokofev1998,Prokofev2004,Boninsegni2006,Hirshberg2019,Hirshberg2020,Myung2022,Feldman2023,RN504,RN505,RN506}.
These zeros will destroy the analytic property of the physical quantities like the free-energy as a function of the 
variable chosen to be continuated when the path goes through them, which underlies the numerical treatments of such analytic continuations.
In this manuscript and the subsequent ones, we address this problem using the idea of LY phase transition theory~\cite{Yang_1952,Lee_1952,Fisher1965,Ouyang_2023}.
We extend the formerly used real parameter $\xi$ into a complex plane, and call the $\xi$s when the partition function equals zero on this plane as ``LY zeros of $\xi$''.
Based on these LY zeroes of $\xi$, we elaborate the FSP from the perspective of analytic continuation for complex functions.
The path-integral representation of quantum mechanics will be used during the discussions, since it gives the most straightforward expansion of the partition
function for the permutation group theory to be utilized so that the exchange symmetry of the many-body system can be addressed rigorously.
However, we note that the conclusions for analytic continuations are not limited to the path-integral expression of the thermodynamic quantities, 
but general to the quantum descriptions of these thermodynamic quantities~\cite{RN511,Xiong2022,Xiong2023,dornheim2024a,RN546}.
In so doing, we can rationalize the successes and challenges of earlier studies in a simple and consistent manner, and provide a completely different perspective 
for later numerical treatments of the FSP based on analytic continuations.
The paper is organized as follows.
In section II, we detailed the derivation of the partition function for indistinguishable particles as a polynomial in $\xi$.
This forms the basis for later theoretical analysis.
Then, the distribution of the LY zeros of $\xi$ at $0$~K is studied in Sec.~\ref{0K}, where we proved that they distribute at $\xi=-1$, $-1/2$, $-1/3$, $\cdots$, $-1/(N-1)$,
regardless of the specific form of the inter-particle interactions and external potentials, as long as the quantum system has bound ground states and discrete energy 
levels in the low-energy region.
The LY zero of $\xi$ at $-1$ is special.
It brings an extra term of the free energy in comparison with the physical systems at other $\xi$, e.g. the bosonic one at $\xi=1$.
Consequently, a phase transition exists if a path is chosen to connect it with a bosonic system with the same potential energies.
In Sec.~\ref{thirdlaw}, we provide discussions of this special point.
In the end, a brief summary of this manuscript is given in Sec.~\ref{Conclusion}.
This paper is the first one of a series composed by three papers, focusing on setting up the theoretical framework for the discussions.
Results at the 0~K limit are also presented.
Finite-$T$ and some numerical results will be summarized in the upcoming ones.
\section{Partition function for indistinguishable particles}
\label{PFa}
In this section, we shall derive a general form of the partition function $Z(N,\beta)$ for indistinguishable $N$-particles as a polynomial in variable $\xi$.
By concluding the emerging techniques regarding paths of intermediate $\xi$-values, we propose a theoretical framework to analyze them by using the Lee-Yang phase transition theory and regarding $\xi$ as a complex variable.
The coefficients of the $\xi$-polynomials are then simplified using the path-integral formalism, characterized by all possible subgroup ring-polymers.
Accordingly, we detailed the standard derivation of recurrence relation in formulating $Z^{(N)}$, with highlight on the $\xi$-terms.
Example cases for $N\le4$ and general cases for abitrary $N$ are given in the later subsections, respectively.

\subsection{The Partition Function as a Polynomial in $\xi$}
Exchange symmetry of indistinguishable particles is the most significant properties of quantum particles, differing from distinguishable particles in classical physics.
According to the Noether theorem, it leads to exchange parity as eigenvalues, as 
\begin{equation}
    \hat{p}_0 |\psi\rangle = \xi |\psi\rangle,
\end{equation}
where the operator $\hat{p}_0$ imposes a single pair exchange, giving the symmetric states $\xi = 1$ and antisymmetric states $\xi = -1$.
The partition function for $N$ particles can be expressed in an explicit form of all possible permutations, as
\begin{equation}
    \label{Z in general}
    \begin{split}
        Z(N,\beta, \xi) &= \mathrm{Tr}\left[ \hat{\rho}\right] = \frac{1}{N!}\sum_{p\in S_N}\sum_{\psi } \xi^{\sigma(p)} \langle p\psi |e^{-\beta \hat H}|\psi\rangle,\\
    \end{split}
\end{equation}
where $\beta=1/(k_\mathrm{B}T)$, $S_N$ is the $N$-order permutation group associated with the indistinguishable particles. 
$S_N$ has $N!$ elements, each denoted by $p$, and $\sigma(p)$ measures the minimal number of pair permutations to restore the identity element~\cite{Dresselhaus_2008}.

Intuitively, one can view $\xi$ as a variable.
It features the exchange symmetry of particles, typically taking real values as $\xi=1$ for bosons, $\xi=-1$ for fermions.
Notably, different paths comprising the intermediate $\xi$-values---such as a path of anyons $\xi=e^{i\theta}$ represented by $\theta=0\to\pi$, or an artificial one that $\xi$ takes real values from -1 to 1---suggest possible connection between bosons and fermions.
Keilmann \textit{et al.} archieved anyons in 1-D optical lattices with fully tuneable exchange statistics, opening the possiblity of transmuting bosons via anyons into fermions~\cite{keilmann2011}.
Beyond conventional idea that anyons must exist in 1-D or 2-D space, Haldane reformulate ``fractional statistics'' in arbitrary dimensions as a generalization of the Pauli principle~\cite{haldane1991}.
Following Haldane exclusion statistics, Batchelor \textit{et al.} reported that the anyonic parameter induced by dynamical coupling constant can not only interpolates between Bose and Fermi statistics, but also triggers statistics for $|\xi|>1$~\cite{batchelor2006}.
However, there are inherent difficulties to realize anyonic statistics in realistic 3-D system, both for experiments and theoretical simulations by PIMD or PIMC methods.
Instead, Xiong \textit{et al.} proposed to use fictitious real interpolating parameter of $\xi=1\to-1$~\cite{Xiong2022}.
Emerging techniques regarding this have been developed to attack FSP by analytically continuating the bosonic results to fermionic systems, leading to both successes and failures.
Dornheim \textit{et al.} used this strategy to explore systems where FSP exists, such as uniform electron gas on large length scales and electrons in warm dense quantum plasmas~\cite{dornheim2024a,dornheim2025}.
The simulation results seem promising but are still not satisfactory, especially for low $T$s~\cite{dornheim2024a,xiong2024}. 
More complicated numerical techniques such as improved extrapolation function are introduced, while the intrinsic physics underlying this artificial methods has never been addressed.

Here, we aims to provide a theoretical framework to analyze above paths systematically.
Utilizing the Lee-Yang phase transition theory, we proposed to regard $\xi$ as a complex variable and investigate the LY zeros of $\xi$.
This is also associated with the Haldane exclusion statistics~\cite{haldane1991}, while we highlight the analytical properties of the partition function when altering $\xi$.
Since $\sigma(p\in S_N)$ ranges from 0 to $N-1$, we shall rewrite Eq.~\eqref{Z in general} in forms of 
\begin{equation}
    \label{polynomial in general}
    Z(N,\beta,\xi) = Z(\xi) = \sum_{j=1}^N c_j \xi^{j-1},
\end{equation}
where the $Z$'s dependency on $N$, $\beta$, or $\xi$ will be omitted in the following text when there is no ambiguity.
The coefficients of the $\xi$-polynomial are given by
\begin{equation}
    \label{coeff in general}
    c_j = c_j(N,\beta) = \frac{1}{N!}\sum_{\psi }\sum_{\sigma(p)=j} \langle p\psi |e^{-\beta \hat H}|\psi\rangle,
\end{equation}
where the second summation is over permutations $p$ satisfying $\sigma(p)=j$.
In the conventional studies of the FSP based on analytic continuations, both $\xi$ and $c_j$ are treated as real numbers.
Following Lee and Yang in Ref.~[\onlinecite{Lee_1952}], we keep the coefficients unchanged and extend $\xi$ to the complex plane.
In so doing, one shall obtain $N-1$ complex roots for the $\xi$-polynomial, which we call ``the LY zeros of $\xi$'', where the partition function equals zero.
When intersecting these zeros, the ensemble average of observables as different orders of derivatives of $\ln Z$ experience non-analytical behaviors and the continuation fails.
This provides a golden rule to tackle the feasibility of intermediate $\xi$-paths for the analytic continuation.

\begin{figure}[b]
    \centering
    \includegraphics[width=\linewidth]{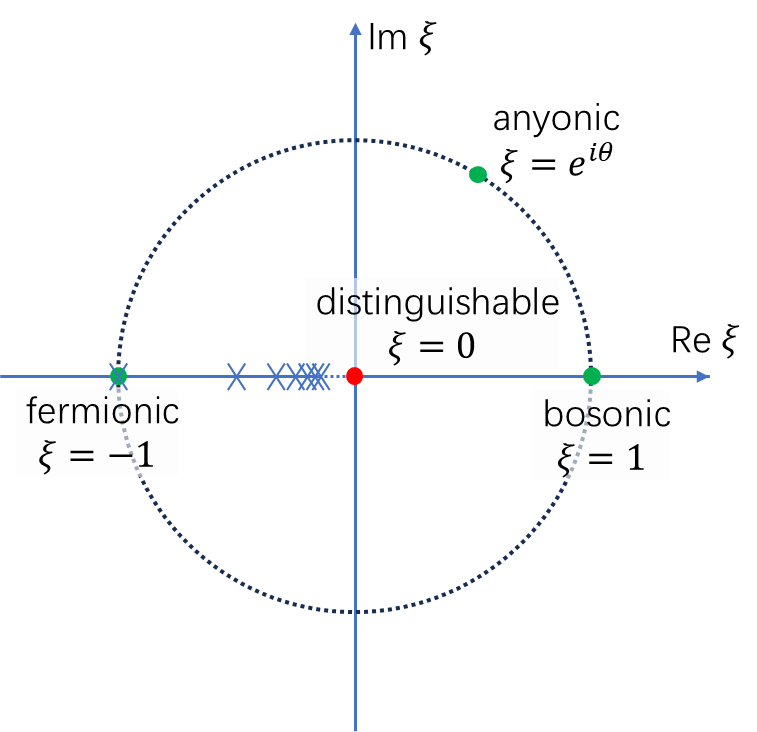}
    \caption{The complex plane of $\xi$. The crosses indicate the exact positions of the LY zeros of $\xi$ at 0~K, which will be proven in section~\ref{0K}. The distinguishable particle systems are characterized by $\xi = 0$, while the indistinguishable particle systems correspond to the unit circle $|\xi|=1$. $\xi=1$ ($\xi=-1$) means the bosonic (fermionic) system, and the other points $\xi = e^{i\theta}$ mean anyonic systems. Using Lee-Yang phase transition theory, we shall consider the LY zeros of $\xi$ in the whole complex plane.}
    \label{complexxi}
\end{figure}

\subsection{The Path-Integral Formalism for the Partition Function}
Notably, Eq.~(\ref{polynomial in general}) and Eq.~(\ref{coeff in general}) is a general form, applicable to arbitrary $\hat{H}$ and being independent of the representations 
of quantum mechanics and the choice of the basis set.
Nevertheless, for practical implementations at finite-$T$s, we choose the path-integral formalism to simplify $c_j$.
The path-integral methods enable transforming the wavefunction problems to classical ring-polymers.
To utilize this convenience, we shall rewrite the partition function in the basis set of the coordinates of the particles, i.e. the coordinate representation of quantum mechanics, as 
\begin{equation}
    \label{anyon}
        Z(N,\beta,\xi) =\frac{1}{N!}\sum_{p\in S_N}\xi^{\sigma(p)}\int d\mathbf{R} \langle p\mathbf{R} |e^{-\beta \hat H}|\mathbf{R}\rangle,
\end{equation}
where $\mathbf{R} = \{R_1,\cdots,R_N\}$ is the set of the coordinates of the particles, and thus $d\mathbf{R}=\prod_{j=1}^{N}dR_j$. 
The Hamiltonian operator is written in a standard form, as
\begin{equation}\label{harmit}
    \hat{H} = \hat{T} + \hat{V},
\end{equation}
where $\hat{T}$ and $\hat{V}$ is the kinetic energy and potential energy operators, respectively.
In the path-integral formalism, the classical $P$-bead ring-polymers serve as an isomorphism of the quantum $N$-particle system, by giving the equivalent ensemble statistics in the limit $P\to \infty$.
Numerical implementations with a finite but proper $P$, being propotional to $\beta$, can also archive converged results.
For simplicity, we omit this limit in the following texts.
Here, each time slice is a replica of the original system, notated as $\mathbf{R}^{(i)} = \{R_1^{(i)},\cdots,R_N^{(i)}\}$, and $\mathbf{R}^P = \{\mathbf{R}^{(1)},\cdots,\mathbf{R}^{(P)}\}$ for coordinates of all the $N\cdot P$ fictitious particles. 
When reducing quantum operators to their classical counterparts, the partition function as the integration over particles is recast into the one over all beads, as
\begin{equation}
    \begin{split}
        \label{partition function in coordinates}
        Z(N,\beta,\xi) &=\int d\mathbf{R}^P~e^{-\beta [K(\mathbf{R}^P) + V(\mathbf{R}^P)]}\\
        &=\frac{1}{N!}\sum_{p\in S_N}\xi^{\sigma(p)}Z_p(N,\beta,\xi),
    \end{split}
\end{equation}
where $d \mathbf{R}^P = \prod_{l=1}^{P} d \mathbf{R}^{(l)} = \prod_{l=1}^{P}\prod_{j=1}^{N} d R_j^{(l)}$, and $Z_p$ represents the path-intergal related to permutation $p$ between indistinguishable particles, as
\begin{equation}
    \label{partition function in permutation}
    Z_p = \int d\mathbf{R} \langle p\mathbf{R} |e^{-\beta \hat H}|\mathbf{R}\rangle = \int d\mathbf{R}^P~e^{-\beta [K_p(\mathbf{R}^P) + V(\mathbf{R}^P)]}.
\end{equation}

Here, we express the $K_p$ and $V$ terms as functions of the ring-polymer coordinates.
The potential operator $\hat{V}$ simply becomes the average of potential energies, as
\begin{equation}
    V(\mathbf{R}^P) = \frac{1}{P} \sum_{l=1}^{P} V(\mathbf{R}^{(l)}),
\end{equation}
where $V(\mathbf{R}^{(l)})$ is the potential energy within each own replica, explaining the absence of $p$-dependency in the $V$-term in Eq.~\eqref{partition function in coordinates}.

The tricks are performed mainly on the kinetic part. 
The kinetic operator $\hat{T}$ is formulated as an additional spring potential terms $K_p(\mathbf{R}^P)$ between neighboring beads.
Different from $V$-term, $K_p(\mathbf{R}^P)$ depends on the ring-polymer configurations which can reflect the effect of permutation $p$.
For the identical permutation $I$, $K_{p=I}$ is written in the same form as for distinguishable particles:
\begin{equation}
    \label{spring term}
    \begin{split}
        K_{p=I}(\mathbf{R}^P) = \sum_{j=1}^N\sum_{l=1}^P \frac{1}{2} m\omega^2_P \left( R_j^{(l)} - R_j^{(l+1)} \right)^2,
    \end{split}
\end{equation}
where $\omega_P = \sqrt{P}/(\beta\hbar)$ is the spring constant of neighboring beads.
In this configuration, the boundary condition is
\begin{equation}
    p=I:~R_j^{(P+1)} = R_j^{(1)},
\end{equation}
where each particle ends its path with itself.

In another case, when there is a permutation between particles, the involved ring-polymer opens and reconnects to form a larger ring-polymer.
Without loss of generality, if there is an exchange $p=(12)$ between the first and second particles, then $K_{p=(12)}$ remains the form of Eq.~\eqref{spring term} but changes the boundary condition to 
\begin{equation}
    p=(12):~R_1^{(P+1)} = R_2^{(1)}, R_2^{(P+1)} = R_1^{(1)}, R_{j\neq1,2}^{(P+1)} = R_j^{(1)}.
\end{equation}
The configurations of these two basic cases are plotted, as shown in Fig.~2(a). 
For more complicated permutations, one should resort to the recurrence relation proposed by Hirshberg in Ref.~[\onlinecite{Hirshberg2019}], as detailed in the next two subsections.

Before elaborating on the mathematical derivations, we note that the partition function can be simplified by considering the structure of permutations.
Since all the particles and the beads are identical, permutation $p$ belonging to the same class $Q$ shares a unique topological structure for their ring-polymer configurations, and they contribute equally to the partition function~\cite{Lyubartsev1993,Feldman2023}.
For instance, the permutations from $123$ to $132$ and $213$ belong to the same class of $(1)(23)$ and correspond to the same topology, featuring one 1-cycle and one 2-cycle ring-polymers.
We denote the permutation class $Q$ as 
\begin{equation}
    Q\equiv 1^{\gamma_1}\cdots N^{\gamma_N},
\end{equation}
where there are $\gamma_m$ instances of $m$-cycle (permuting all the $m$ elements in a single cycle) for $p\in Q$, with $\sum_{m=1}^{N} m\cdot \gamma_m = N$.
With this, Eq.~\eqref{anyon} can be reduced to a sum over unique configurations, i.e. over the classes $Q$s of permutation group, as
\begin{equation}
    \label{anyon1}
    Z(\xi)=\frac{1}{N!}\sum_{Q}\sum_{p\in Q}\xi^{\sigma(p)}Z_p=\sum_{Q}\xi^{\sigma(Q)}\frac{F_Q}{N!}Z_Q,
\end{equation}
where $\sigma(Q) = \sum_{m=1}^{N} (m-1)\gamma_m$. 
$F_Q$ is the number of elements for $p\in Q$ in $S_N$, given by Ref.~[\onlinecite{Dresselhaus_2008}], as 
\begin{equation}
\label{FQN}
F_Q=\frac{N!}{\big(1^{\gamma_1}\gamma_1!\big)\big(2^{\gamma_2}\gamma_2!\big)\cdots\big(N^{\gamma_N}\gamma_N!\big)}.
\end{equation}
Comparing to Eq.~\eqref{Z in general} and Eq.~\eqref{coeff in general}, we rewrite $c_j$ as 
\begin{equation}
    \label{coeff in Q}
    c_j = \frac{1}{N!}\sum_{\sigma(Q)=j-1} F_QZ_Q.
\end{equation}
\begin{figure}[b]
    \centering
    \includegraphics[width=\linewidth]{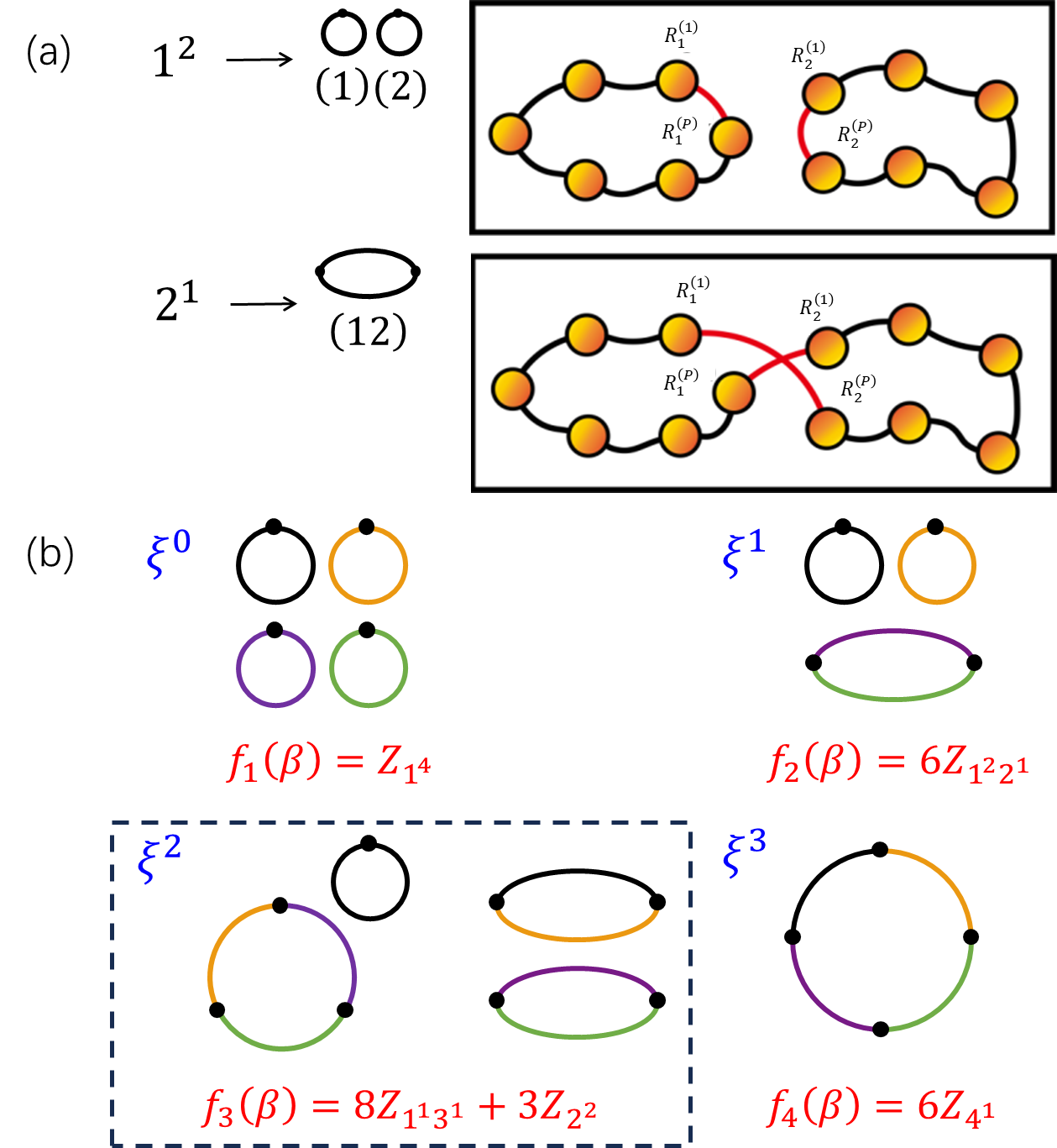}
    \caption{(a) The schematic of the ring-polymer configurations for identical permutation (also for distinguishable particles) and single pair permutation in the path-integral formalism, respectively. The schematic notations for configurations are following Ref.~[\onlinecite{Hirshberg2019}]. (b) The schematic representation of $\xi$-term of the partition function, organized by $\xi$-orders. The 4-particle system are taken here as an example: the 24 elements of its permutation group belong to 5 classes and their representive ring-polymer configurations are presented, corresponding to $Z_{1^4}$, $Z_{{1^2}{2^1}}$, $Z_{{1^1}{3^1}}$, $Z_{2^2}$, and $Z_{4^1}$. The coefficients of $\xi$-terms might come from distinct ring-polymer configurations. For example, $f_3(\beta)$ contains contributions of $Z_{{1^1}{3^1}}$ and $Z_{2^2}$, as indicated by the dashed square.}
    \label{smpc4}
\end{figure}

\subsection{Example cases: $N\le 4$}
Compared to summing over all permutations, the computational cost of Eq.~\eqref{anyon1} has been significantly reduced, but it still scales exponentially with the system size.
In the following, we seek practical implementations of this expression.
We start with the simplest cases to gain insights.
With the above notations, we can write out the partition functions for $N=1$ to $4$ in alignment of $\xi$-orders, as
\begin{align}
    \label{Z1}
    Z(N=1,\xi)=&Z_{1^1} \xi^0,\\
        \label{Z2}
    Z(N=2,\xi)=&\frac{1}{2}(Z_{1^2}\xi^0+Z_{2^1} \xi^1 ),\\
        \label{Z3}
    Z(N=3,\xi)=&\frac{1}{6}(Z_{1^3}\xi^0+3Z_{1^12^1}\xi^1 +2Z_{3^1}\xi^2),\\
        \label{Z4}
    Z(N=4,\xi)=&\frac{1}{24}\big[Z_{1^4}\xi^0+6Z_{1^2 2^1}\xi^1+\big. \notag\\
    &\big.(8Z_{1^13^1}+3Z_{2^2})\xi^2+6Z_{4^1}\xi^3\big].
\end{align}
It is clear that different $Z_Q$s comprise the coefficients of $\xi$-polynomial.
In Fig.~\ref{smpc4}(b), we show the topological ring-polymer structure of the 4-particle system as an example, also arranged by $\xi$-order terms.
In the terms of partition function $Z(N+1, \xi)$, $(N+1)$-cycle emerges as a new term, while the other $j$-cycle terms ($j\le N$) have appeared in prior $Z(j)$s expressions ($j\le N$).
The terms in $Z(N=4,\beta,\xi)$ can be decomposed into combinations of different $Z_Q$s that appear in the cases for $N=1,2,3$.
To fully utilize the information of known partition functions for $j<N$ and further reduce computational costs, Hirshberg \textit{et al.} proposed performing path-integral calculations of $N$ by using the recurrence relation~\cite{Hirshberg2019,Hirshberg2020}.

\subsection{Recurrence Relation for General Cases}
The recurrence relation technique establishes a practical manner for simulating indistinguishable particles.
In this subsection, we detail its derivation and obtain a similar recurrence relation for the partition function.
To explicitly illustrate the connection between systems of different sizes, we denote the $N$-dependency of variables with a superscript $[i,j]$, indicating a subsystem composed by particles from index $i$ to $j$.
The partition function for an $N$-particle system, as given in Eq.~\eqref{partition function in coordinates}, is rewritten as
\begin{equation}
\label{partition-pi}
Z(N) \equiv Z^{[1,N]}=\int d\mathbf{R}^P e^{-\beta (K^{[1,N]}+V^{[1,N]})},
\end{equation}
For simplicity, we denote the weighting coefficients contributed by $K^{[1,N]}$ and $V^{[1,N]}$ as
\begin{equation}
    \begin{split}
        W^{[1,N]} = e^{-\beta K^{[1,N]}},
    \end{split}
\end{equation}
and 
\begin{equation}
    U^{[1,N]} = e^{-\beta V^{[1,N]}},
\end{equation}
respectively. Thus, the partition function becomes 
\begin{equation}
    Z^{[1,N]}=\int d\mathbf{R}^P W^{[1,N]} U^{[1,N]}.
\end{equation}
The spring energy term $W$ can be expressed in a form similar to Eq.~\eqref{anyon1}, as 
\begin{equation}
    \label{W full}
    W^{[1,N]}=\sum_{Q}\xi^{\sigma(Q)}\frac{F_{Q}}{N!}W_Q^{[1,N]}.
\end{equation}
For $W_Q$ under a specific permutation $p\in Q$, we note that particles interact only when they are permuted and connected within a larger ring-polymer.
Otherwise, there is no spring interaction between them.
Based on this, we can decompose $W_Q$ into a product of a series of $W$s for the individual cycles in the ring-polymer configuration corresponding to its permutation structure,
\begin{equation}
\label{T_Q}
W_Q\equiv W_{1^{\gamma_1}2^{\gamma_2}\cdots N^{\gamma_N}}=\prod_{i=1}^{N} \left[W_{i^1}\right]^{\gamma_i}.
\end{equation} 
Furthermore, it follows that 
\begin{equation}
    W_Q^{[1,N]} = W_{Q^\prime}^{[1,N-k]}W_{k^1}^{[N-k+1,N]},
\end{equation}
where $Q^{\prime}\equiv 1^{\gamma_1}\cdots k^{\gamma_k-1}\cdots N^{\gamma_N}$ with only one $k$-cycle permutation is taken out from $Q$.
Substituting this into Eq.~\eqref{W full}, we find that $W^{[1,N]}$ can be expressed in terms of those for smaller systems, through 
\begin{equation}
\label{recu11}
W^{[1,N]}=\frac{1}{N}\sum_{k=1}^N\xi^{k-1}W^{[1,N-k]}W_{k^1}^{[N-k+1,N]}.
\end{equation}
This is the so-called ``recurrence relation'', as proposed by Hirshberg \textit{et al.} in Refs.~[\onlinecite{Hirshberg2019}] and [\onlinecite{Hirshberg2020}].
Especially, in the $k=N$ term, $W^{[1,0]}=e^{-\beta K^{[1,0]}}$ means there is no particle in this loop, thus $K^{[1,0]}=0$ and $W^{[1,0]}=1$.

\begin{figure}[b]
    \centering
    \includegraphics[width=\linewidth]{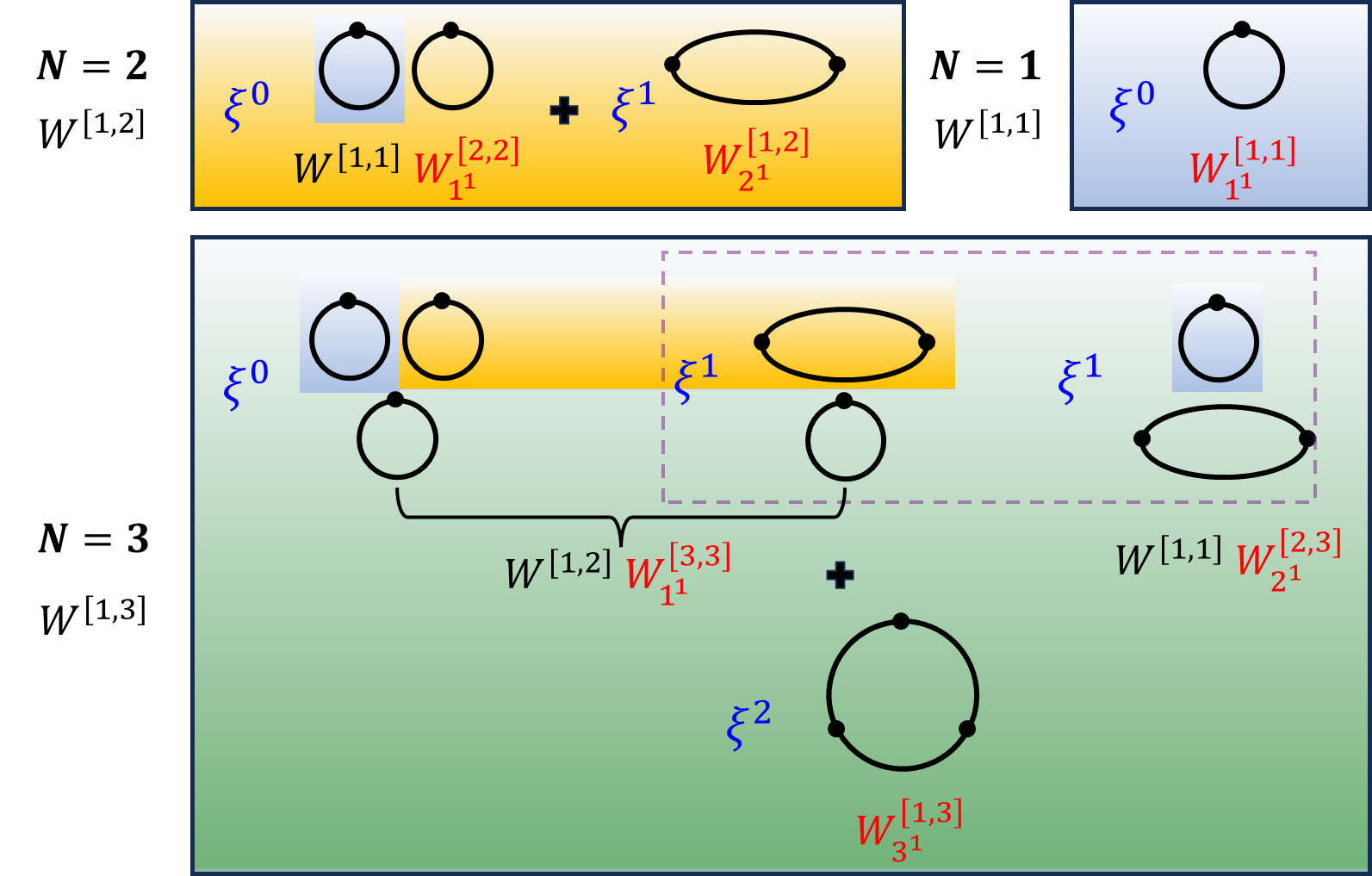}
    \caption{Schematic of recurrence relation in alignment of $\xi$-orders. The $N=3$ system (in green) have contributions from different combinations: $N=1$ term (in cyan) $Z^{[1,1]}$ plus the 2-cycle term $Z_{2^1}$, $N=2$ term (in yellow) plus the 1-cycle term $Z_{1^1}$, and the new 3-cycle term $Z_{3^1}$. Due to the combinations, there can be mixed terms contributing to $\xi$-orders. For example, $\xi^1$ terms, as labeled by the box with dashed lines, come from both $W^[1,2]W_{1^1}^{[3,3]}$ and $W^[1,1]W_{2^1}^{[2,3]}$. }
    \label{combine}
\end{figure}

In Fig.~\ref{combine}, we use a 3-particle example to explain how this scheme works in practice, with a special focus on $\xi$-terms.
It should be noted that $W^{[1,N-k]}$ is also a $\xi$-polynomial of $N-k$ order.
Therefore, there are crossovers between different terms in the recurrence relation.
For example, the $W^{[1,2]}$ terms contribute to both $\xi^0$ and $\xi^1$, as shown in Fig.~\ref{combine}.
Eq.~\eqref{recu11} also enables evaluating effective force by a similar recurrence relation, as weighted average force due to all ring-polymer configurations, and further enables sampling ring-polymer configurations.
Here, we focus on the partition function as the $\xi$-polynomial.
Inspired by above treatments for $W$-terms, we shall also express $Z(\xi)$ in forms of a recurrence relation.
Regarding that $U^{[1,N]}$ contains interacting terms, we introduced a circle sign $\circ$ to denote the ``multiplication'' between partition functions of different subsystems, as
\begin{equation}
\begin{split}
Z_Q &=Z_{Q_1}\circ Z_{Q_2} \\
&\equiv\lim_{P\to \infty}\int  d\mathbf{R}^PW_{Q_1}^{[1,N_1]}W_{Q_2}^{[N_1+1,N_1+N_2]}U^{[1,N_1+N_2]}.\raisetag{-10pt}
\end{split}
\end{equation}
This multiplication means: when adding the subsystem of size $N_2$ to the one of $N_1$, the kinetic $W$-terms in $Q_1$ and $Q_2$ are isolated ring-polymers, while the potential $U$-terms have interactions.
Replacing $Z_Q$ by $Z_{1^{\gamma_1}2^{\gamma_2}\cdots N^{\gamma_N}}$, one further has
\begin{equation}
Z_{\prod_i i^{\gamma_m}}\circ Z_{\prod_i i^{\gamma_n}}=Z_{\prod_i i^{\gamma_m+\gamma_n}}.
\end{equation}
Therefore, the total partition function of the $N$-particle system can also be given by a recurrence relation, as
\begin{equation}
    \label{recursive}
    Z^{[1,N]}=\frac{1}{N}\sum_{n=1}^N\xi^{n-1}Z^{[1,N-n]}\circ Z_{n^1}, 
\end{equation}
where $Z^{[1,0]}\equiv 1$.
The mathematical structure of the operations defined by $\circ$ is similar to the multiplication defined on positive real numbers, where the set of elements are limited to 
only positive integers.
In this case, the rules of commutation, association and distribution are satisfied.
But for each element of the group, its inverse with respect to the identity element does not exist.
One special situation of the expression in Eq.~\eqref{recursive} is for the systems without interactions, where $\circ$ just becomes multiplication of numbers, 
as we have shown for the kinetic part of the partition function in Eq.~\eqref{T_Q}.
\section{LY zeros of $\xi$ at 0~K}
\label{0K}
The expression in Eq.~\eqref{recursive} offers a clear analytical form of the partition function of the $N$-particle.
With this, we are able to investigate the distribution of LY zeros of $\xi$ on the complex plane at any temperature.
Notably, the result for the 0~K limit is elegant: the LY zeros of $\xi$ are located at $-1$, $-1/2$, $-1/3$, $\cdots$, $-1/(N-1)$.
While the result for finite $T$s will be addressed in the next papers of our series.
The original formula as Eq.~\eqref{anyon1} can be simplified significantly for the 0~K limit, by noting that the $Z_Q$s corresponding to different permutation classes of $S_N$ are equal to each other at 0~K.
We will prove the equality of the $Z_Q$s in the first subsection.
This property persists whenever the system has bounded ground states and discrete energy levels in the low-energy region, which is robust in realistic systems.
Then in Sec.~IIIB , using this property, we will show the distribution of LY zeros of $\xi$ at 0~K as a nontrivial theorem.
These can help us to understand the successes/failures of existing studies of the FSP based on analytic continuations in a consistent manner.

\subsection{The Equality of the $Z_{Q}$s at 0~K}
\label{equality}
Intuitively, one can imagine that at 0~K, the spring interactions between neighboring beads are infinitely weak.
As such, using different ring-polymers as shown in Fig.~\ref{smpc4} does not make much difference for the numerical sampling over the coordinate space
in Eq.~\eqref{partition function in permutation}.
Therefore, there is a large probability for the different $Z_Q$s to contribute equally to the final $Z^{(N)}$, subjected to the fact that these contributions may
cancel each other due to the different signs of $\xi^{n-1}$ in Eq.~\eqref{anyon1}.
The proof of this property, however, can not rely solely on this physical intuition.
Our strategy is to use the product states of the many-body system, other than the determinant states or permanent states, as the basis set to express the partition functions
and demonstrate using mathematics that these terms equal.
The use of this basis set is equivalent to the use of the coordinate space to express $Z^{(N)}$ and $Z_Q$ in Eqs.~\eqref{partition-pi} and \eqref{partition function in permutation}.
The influences of the permutation operations, as shown in Eq.~\eqref{anyon1}, are addressed by the term of $\xi^{\sigma(Q)}$ on the right-hand-side of it.
Thus, when talking about $Z_Q$ using ring-polymer, it is already assumed that the particles are distinguishable, with exchange taken care by the topology of the 
ring-polymer and the corresponding $\xi^{\sigma(Q)}$.
The ground state of the quantum system should be discussed in the Hilbert space composed by the product states.
Based on this, we will prove that the partition function of $Z_{1^N}$, i.e. the one corresponding to the identity element of the permutation group, equals the partition 
function of $Z_{Q}$ for any class of the permutation group at 0~K.
This property will allow us to simplify the expression of the polynomial in Eq.~(\ref{anyon1}).
We start from the partition function of $Z_{1^N}$.
Here, the permutation operator $\hat{\sigma}$ is the identity element $\hat{\sigma}_0$.
Therefore, one has
\begin{equation}
\begin{split}
Z_{1^n}\left(\beta\right)&=\text{Tr}\left[\hat{\sigma}\hat{\rho}\right]\\
&=\int d\mathbf{R} \rho\left(R_{1},\cdots,R_{N};R_1,\cdots,R_N\right)\\
&=\sum_ie^{-\beta\epsilon_i}\left<\psi_i\middle|\hat{\sigma}_0\middle|\psi_i\right>\\
&=\sum_ie^{-\beta\epsilon_i}.
\end{split}
\end{equation}
The partition functions of other permutations with the number of single cycle being $n_1$, on the other hand, reads
\begin{equation}
\begin{split}
Z_{1^{n_1}\cdots}&=\text{Tr}\left[\hat{\sigma}\hat{\rho}\right]\\
&=\int d\mathbf{R}
\rho\left(R_{\sigma_1},\cdots,R_{\sigma_N};R_1,\cdots,R_N\right)\\
&=\sum_ie^{-\beta\epsilon_i}
\left<\psi_i\middle|\hat{\sigma}\middle|\psi_i\right>.
\end{split}
\end{equation}
Here, $\epsilon_i$ and $\psi_i$ are the eigenenergy and eigenvector of the many-body system in the Hilbert space composed by the product states.
Sum over them is mathematically equivalent to the integral over the coordinate space in Eq.~\eqref{partition function in permutation}.
The ratio between $Z_{1^N}$ and $Z_{1^{n_1}\cdots}$ is
\begin{equation}\label{zratio}
\begin{split}
\frac{Z_{1^{n_1}\cdots}}{Z_{1^n}}&=\frac{\sum_ie^{-\beta\epsilon_i}
\left<\psi_i\middle|\hat{\sigma}\middle|\psi_i\right>}
{\sum_ie^{-\beta\epsilon_i}}\\
&=\frac{\sum_{i=0}^{g-1}\left<\psi_i\middle|\hat{\sigma}\middle|\psi_i\right>
+\sum_{i=g}^\infty e^{-\beta\left(\epsilon_i-\epsilon_0\right)}
\left<\psi_i\middle|\hat{\sigma}\middle|\psi_i\right>}
{g+\sum_{i=g}^\infty e^{-\beta\left(\epsilon_i-\epsilon_0\right)}},\\\raisetag{0pt}
\end{split}
\end{equation}
where $g$ is the degeneracy of the ground state.
At the $T\to0$ limit, the exponential factor $e^{-\beta\left(\epsilon_i-\epsilon_0\right)}$ tends to be zero, while other numbers keep constant, so the limit of the ratios are
\begin{equation}
\lim_{T\to0}\frac{Z_{1^{n_1}\cdots}}{Z_{1^n}}=\frac{1}{g}
\sum_{i=0}^{g-1}\left<\psi_i\middle|\hat{\sigma}\middle|\psi_i\right>.
\end{equation}
To prove that this limit is 1, it is sufficient to prove that the ground states are all invariant to permutations and thus $\left<\psi_i\middle|\hat{\sigma}\middle|\psi_i\right>=1$ for $i=0,\cdots,g-1$.
To achieve this goal, we follow the route provided by Lieb and Seiringer in Ref.~[\onlinecite{Lieb2010}], and prove it in two steps.
A slight difference is that in the first step we start from the positivity of the density matrix elements
\begin{equation}
\rho\left(\mathbf{R}^{\prime};\mathbf{R}\right)=\frac{\cdots}{\cdots}
\int_{\mathbf{R}\left(0\right)=\mathbf{R}}^{\mathbf{R}\left(\beta\right)=\mathbf{R}^{\prime}}
D\left[\cdots\right]e^{-S\left[\mathbf{R}(\tau)\right]}>0.
\end{equation}
This positivity exists because in imaginary time $\tau$ the Euclidean action is real and therefore contributions of all paths are positive.
For a vector $\Psi$ with unit modular in the Hilbert space composed by the product states $\left|\psi_i\right>$, we have
\begin{equation}
\begin{split}
\sum_i\left|\left<\psi_i\middle|\Psi\right>\right|^2&e^{-\beta\epsilon_i}=
\left<\Psi\middle|\hat{\rho}\middle|\Psi\right>\\
&=\int d\mathbf{R}^\prime d\mathbf{R}
\rho\left(\mathbf{R}^{\prime};\mathbf{R}\right)\Psi^*\left(\mathbf{R}^{\prime}\right)\Psi\left(\mathbf{R}\right).
\end{split}
\end{equation}
Now, consider $\Phi\left(\mathbf{R}\right)=
\left|\Psi\left(\mathbf{R}\right)\right|$, which also has unit modular, one has
\begin{equation}\label{inequal}
\begin{split}
\left<\Phi\middle|\hat{\rho}\middle|\Phi\right>&=\int d\mathbf{R}^{\prime}d\mathbf{R}\rho\left(\mathbf{R}^{\prime};\mathbf{R}\right)
\left|\Psi^*\left(\mathbf{R}^{\prime}\right)\Psi\left(\mathbf{R}\right)\right|\\
&\ge\left<\Psi\middle|\hat{\rho}\middle|\Psi\right>.
\end{split}
\end{equation}
Then, we look at the case when $\Psi$ is a ground state of many-body system, where Eq.~\eqref{inequal} still holds.
This ground state has the property that
$\left<\Psi\middle|\hat{\rho}\middle|\Psi\right>=e^{-\beta\epsilon_0}\ge\left<\Phi\middle|\hat{\rho}\middle|\Phi\right>$, since this $\Phi\left(\mathbf{R}\right)$ is not
necessarily the ground state.
Combining these two inequalities, one gets the conclusion that when $\Psi$ is a ground state, $\left<\Psi\middle|\hat{\rho}\middle|\Psi\right>=\left<\Phi\middle|\hat{\rho}\middle|\Phi\right>$.
Since $\Phi\left(\mathbf{R}\right)=\left|\Psi\left(\mathbf{R}\right)\right|$, the equality between 
$\left<\Psi\middle|\hat{\rho}\middle|\Psi\right>$ and $\left<\Phi\middle|\hat{\rho}\middle|\Phi\right>$ holds only when $\Psi=e^{i\theta}\left|\Psi\right|$, 
where $\theta$ is a parameter independent on the variable of the vector $\Psi$.
This constitutes the conclusion for the first step of our proof.
In the second step, we prove that the restriction is actually even stronger, and $\left<\psi_i\middle|\hat{\sigma}\middle|\psi_i\right>=1$ holds for 
the ground states.
To verify this, we introduce a projector,
\begin{equation}
\hat{P}=\frac{1}{N!}\sum_{\sigma\in S_n}\hat{\sigma},
\end{equation}
which can extract the part that belong to the trivial representation from the state as $\left|\Psi_s\right>\equiv\hat{P}\left|\Psi\right>$ and we denote the rest as $\left|\Psi_a\right>\equiv\left|\Psi\right>-\left|\Psi_s\right>$.
According to the theory of irreducible representations, we have
\begin{equation}
\left<\Psi_s\middle|\Psi_a\right>=
\left<\Psi_s\middle|\hat{H}\middle|\Psi_a\right>=0,\quad
\hat{P}\left|\Psi_a\right>=0.
\end{equation}
The latter equation gives a condition for the wave function of the rest part,
\begin{equation}
\sum_{\sigma\in S_n}\Psi_a\left(R_{\sigma_1},\cdots,R_{\sigma_n}\right)=0,
\end{equation}
meaning that $\Psi_a$ is either identically zero or changes signs upon permutation.
Then, as long as we can prove that $\Psi_a$ does not change signs upon permutation, we will get the conclusion that $\Psi_a=0$ and $\Psi=\Psi_s$, i.e. the ground state is 
invariant with respect to permutations, and complete the proof.
To see this, we look at the constraint on energy, where we have
\begin{equation}
\begin{split}
\epsilon_0&=\left<\Psi\middle|\hat{H}\middle|\Psi\right>
=\left<\Psi_s\middle|\hat{H}\middle|\Psi_s\right>
+\left<\Psi_a\middle|\hat{H}\middle|\Psi_a\right>\\
&\ge\epsilon_0\left(
\left<\Psi_s\middle|\Psi_s\right>+\left<\Psi_a\middle|\Psi_a\right>\right)=\epsilon_0 \left<\Psi \middle|\Psi \right>
=\epsilon_0.
\end{split}
\end{equation}
The equality holds only when both $\Psi_s$ and $\Psi_a$ are ground states.
For the ground state, our first step gives us the conclusion that it has the form of $e^{i\theta}\left|\Psi\right|$, meaning that it does not change the signs upon permutation.
Therefore, for $\Psi_a$, it must be zero.
This leads to the conclusion that being a ground state, $\left|\Psi\right>$ must be permutationally invariant, hence $\left<\psi_i\middle|\hat{\sigma}\middle|\psi_i\right>=1$, 
which completes our proof.
\subsection{Distribution of the LY zeros of $\xi$ at 0~K}
\label{Zero}
Now we focus on the partition function and the LY zeros. 
Since the partition function of every permutation equals to each other, as proven in Sec.~IIIA, $Z^{(N)}$ in Eqs.~\eqref{Z1}-\eqref{Z4} at $0~K$ can be recast into
\begin{equation}
    \label{reduced123}
    \begin{split}
    Z(N=1)&=Z_{1^1} ,\\
    Z(N=2)&=\frac{1}{2}(1+\xi)Z_{2^1},\\
    Z(N=3)&=\frac{1}{6}(1+\xi)(1+2\xi)Z_{3^1},\\
    Z(N=4)&=\frac{1}{24}(1+\xi)(1+2\xi)(1+3\xi)Z_{4^1}.
    \end{split}
\end{equation}
From these, we expect that the partition function of a $N$-particle system can be expanded as: 
\begin{equation}
    \label{reducedn}
    Z(N,\beta\to\infty,\xi)=\frac{1}{N!}\prod_{n=0}^{N-1}(1+n\xi)Z_{N^1}.
\end{equation}
In the following, we will prove that this equation holds in a mathematically rigorous manner.
The trick is to assume that the Eq.~\eqref{reducedn} holds for all $n \le N$.
Then, as long as we can use the recurrence relation to prove
\begin{equation}
  \label{reducedn+1_0}
  Z(N+1,\xi)=\frac{1}{(N+1)!}\prod_{n=0}^{N}(1+n\xi)Z_{{(N+1)}^1}.
\end{equation}
Since Eq.~\eqref{reduced123} already permits the earliest extrapolation, Eq.~\eqref{reducedn} will hold for the $N$-particle system rigorously.
With this strategy clear, what we need to do is to put Eq.~\eqref{reducedn} into Eq.~\eqref{recursive}.
This yields
\begin{equation}
\begin{split}
    \label{reducedn+1_01}
    Z(N+1,&\xi)=\frac{1}{N+1}\Big[\xi^N Z_{(N+1)^1}+\\
&\sum_{n=1}^{N}\big(\frac{1}{n!}\xi^{N-n}\prod_{j=0}^{n-1}(1+j\xi)\big)Z_{n^{1}(N-n+1)^1}\Big].
\end{split}
\end{equation}
By using the equality of the different $Z_Q$s, Eq.~\eqref{reducedn+1_01} can be rewritten as, 
\begin{equation}
\begin{split}
    \label{reducedn+1_1}
        Z(N+1,\xi)&= \frac{1}{N+1} \Xi Z_{(N+1)^1} \\
        &=\frac{1}{N+1}\Big[\xi^N\sum_{n=1}^{N}\big(\frac{1}{n!}\xi^{N-n}\prod_{j=0}^{n-1}(1+j\xi)\big)\Big]Z_{{(N+1)}^1},
\end{split}
\end{equation}
where the $\xi$-dependency is extracted as $\Xi$.
Adding the first term in the summation to $\xi^N$ in the coefficient of Eq.~\eqref{reducedn+1_1} yields
\begin{align}
\Xi\equiv&\xi^N+\sum_{n=1}^{N}\big(\frac{1}{n!}\xi^{N-n}\prod_{j=0}^{n-1}(1+j\xi)\big)\nonumber\\
=&(1+\xi)\Big[\xi^{N-1}+\sum_{n=2}^{N}\frac{\xi^{N-n}\prod_{j=2}^{n-1}(1+j\xi)}{n!}\Big].
\end{align}
Then adding the next term in the summation to $\xi^{N-1}$ yields
\begin{align}
\Xi=(1+\xi)(1+2\xi)\Big[\frac{\xi^{N-2}}{2}+\sum_{n=3}^{N}\frac{\xi^{N-n}\prod_{j=3}^{n-1}(1+j\xi)}{n!}\Big].
\end{align}
Repeating the process for $k$ times, we obtain
\begin{align}
    \label{reducedn+1k}
\Xi=\prod_{l=0}^k(1+l\xi)\Big[\frac{\xi^{N-k}}{k!}+\sum_{n=k+1}^{N}\frac{\xi^{N-n}\prod_{j=k+1}^{n-1}(1+j\xi)}{n!}\Big].
\end{align}
Thus, when $k=N-1$, we finally obtain
\begin{equation}
\begin{split}
    \label{reducedn+1}
\Xi=&\xi^N+\sum_{n=1}^{N}\big(\frac{1}{n!}\xi^{N-n}\prod_{j=0}^{n-1}(1+j\xi)\big)\\
=&\prod_{j=0}^{N-1}(1+j\xi)\Big[\frac{\xi}{(N-1)!}+\frac{1}{N!}\Big]=\frac{1}{N!}\prod^{N}_{j=0}(1+j\xi),\raisetag{-20pt}
\end{split}
\end{equation}
Substitution of Eq.~\eqref{reducedn+1} into Eq.~\eqref{reducedn+1_1} yields Eq.~\eqref{reducedn+1_0}, hence the claim in Eq.~\eqref{reducedn} is proven.
With the partition function expressed in Eq.~\eqref{reducedn}, the LY zeros of the $N$-particle quantum system at 0~K are 
\begin{align}
    \label{0zeros}
    &z_1=-1,\ z_2=-\frac{1}{2},\ z_3=-\frac{1}{3},\ \cdots,\nonumber\\
    &z_{N-1}=-\frac{1}{N-1}.
\end{align}
It is obvious that all the LY zeros at 0~K are on the negative real axis of the complex field. 
The position of the first zero $z_1$ means the partition function of fermions equals zero at 0~K.
\section{The LY zero of $\xi$ at $-1$ and the Speciality of the Fermionic Systems}
\label{thirdlaw}
From the above discussions, we know that $z_1=-1$ is a LY zero of $\xi$.
Some people may have the impression that LY zeros bring singularity to the thermodynamic quantities.
Rigorously speaking, this statement is incorrect.
When close enough to the real axis, saying that LY zero breaks the analytic continuity of the thermodynamic quantity as a function of the thermodynamic state function 
might be more appropriate~\cite{Yang_1952,Lee_1952}.
Considering the fact that $\xi=-1$ corresponds to the fermionic system, to avoid possible misunderstanding which might arise from such nuanced interpretations, we provide 
a systematic discussion on the influence of the LY zero of $\xi$ at $-1$ on the behaviors of thermodynamic quantities such as the free energy for the fermionic and bosonic systems.
Our conclusion is that the LY zero of $\xi$ at $-1$ makes the fermionic systems special, with its analytic form of the free energy being different
from the bosonic one and quantum systems composed by other anyons.
This can be deduced from the third law of thermodynamics.
In the following, we go through this procedure, starting from the polynomial expression of the partition function.
With the roots of $Z(\beta,\xi)=0$ labelled by $\{z_i\}$ at 0~K, the polynomial partition function 
of Eq.~\eqref{anyon1} can be written as 
\begin{equation}
\label{pfz0}
Z(\beta,\xi)=\frac{Z_{N^1}}{N!}\prod_{i=1}^{N-1}(\xi-z_i).
\end{equation}
Putting it into the equation of the free energy,
\begin{equation}
\label{FreeEnergy}
F=-k_BT\ln Z(T,\xi)
\end{equation}
we will get
\begin{equation}
\label{freeenergy}
F(T,\pm 1)=-k_BT(\ln Z_{N^1}+\sum_{i=1}^{N-1}\ln |z_i\mp 1| -\ln N),
\end{equation}
for the bosonic and fermionic systems.
According to the third law of thermodynamics, the derivative of the free energy with respect to the temperature equals $0$ at 0~K, i.e. $\beta\to\infty$.
This means that 
\begin{equation}
\label{dFdT}
\lim_{T\to0}  \left[k_B\ln Z(\beta,\pm1)+ 
 k_BT\left(\frac{d\ln Z_{N^1}}{dT}+\sum_{i=1}^{N-1}\frac{dz_i/dT}{z_i\mp1}\right)\right]=0.
\end{equation}
To decipher the analytic form of the third term on the left-hand-side of Eq.~\eqref{dFdT}, we resort to the following polynomial
\begin{equation}\label{ftpoly}
f(T, \xi)\equiv\prod_{i=1}^{N-1}\left(1-\xi/z_i\right).
\end{equation}
According to Eq.~\eqref{pfz0}, it also equals $N!Z(\beta,\xi)/Z_{1^N}$.
From Eqs.~\eqref{anyon1} to \eqref{coeff in Q}, one can see that all the coefficients of the polynomial in Eq.~\eqref{ftpoly} are combinations of $Z_Q/Z_{1^N}$.
Therefore, one can extract a relation between the zeros and the temperature from Eq.~\eqref{ftpoly}, characterized by the total derivative,
\begin{equation}\label{eq:dftotal}
df\left(T, z_i\right)=\left.\frac{\partial f}{\partial T}\right|_{\xi=z_i}dT
+\left.\frac{\partial f}{\partial \xi}\right|_{T}dz_i=0.
\end{equation}
This gives an explicit formula
\begin{equation}\label{dzi}
\frac{dz_i}{dT}=
-\frac{\left.\partial f/\partial T\right|_{\xi=z_i}}{\left.\partial f/\partial \xi\right|_{T}}.
\end{equation}
At the 0~K limit, $f\left(T,\xi\right)$ tends to $\prod_{m=1}^{k-1}(1+m\xi)$, which has fixed non-zero partial derivatives with respect to $\xi$ at the zeros.
Therefore, the denominator of Eq.~\eqref{dzi} has a non-zero limit.
Then, we look at the numerator of Eq.~\eqref{dzi}.
We can use its explicit formula given by Eqs.~\eqref{anyon1} to \eqref{coeff in Q}, i.e.
\begin{equation}\label{eq:pfpT}
\begin{split}
\frac{\partial f}{\partial T}&=\frac{\partial}{\partial T}
\frac{\sum_Q\xi^{\sigma(Q)}F_{N,Q}Z_Q}{Z_{1^N}}\\&=
\sum_Q\frac{-\xi^{\sigma(Q)}F_{N,Q}}{k_BT^2}
\left(\frac{\partial Z_Q/\partial \beta}{Z_{1^N}}
-\frac{Z_Q}{Z_{1^N}}\frac{\partial Z_{1^N}/\partial \beta}{Z_{1^N}}\right)\raisetag{-20pt}
\end{split}
\end{equation}
and expand Eq.~\eqref{zratio} around $\beta=\infty$ to get an estimation of the ratio,
\begin{equation}
	\begin{split}\label{zqz1N1}
		\frac{Z_Q}{Z_{1^N}}
		&=\frac{g+\sum_{i=g}^\infty e^{-\beta\left(\epsilon_i-\epsilon_0\right)}
			\left<\psi_i\middle|\hat{\sigma}\middle|\psi_i\right>}
		{g+\sum_{i=g}^\infty e^{-\beta\left(\epsilon_i-\epsilon_0\right)}}\\&=
		\frac{g+e^{-\beta\left(\epsilon_1-\epsilon_0\right)}\sum_{i=g}^{g+g_1-1}
			\left<\psi_i\middle|\hat{\sigma}\middle|\psi_i\right>
			+o\left(e^{-\beta\left(\epsilon_1-\epsilon_0\right)}\right)}
		{g+g_1e^{-\beta\left(\epsilon_1-\epsilon_0\right)}+o\left(e^{-\beta\left(\epsilon_1-\epsilon_0\right)}\right)}\\&=
		1+\frac{e^{-\beta\left(\epsilon_1-\epsilon_0\right)}}{g}\sum_{i=g}^{g+g_1-1}
		\left<\psi_i\middle|\hat{\sigma}-1\middle|\psi_i\right>
		+o\left(e^{-\beta\left(\epsilon_1-\epsilon_0\right)}\right).\raisetag{0pt}
	\end{split}
\end{equation}
Here, $g_1$ and $\epsilon_1$ are the degeneracy and energy of the lowest excited state.
Similarly, for the derivatives we have
\begin{equation}\label{dzratio}
\begin{split}
\frac{-\partial Z_Q/\partial \beta}{Z_{1^N}}
=&\frac{\sum_i\epsilon_ie^{-\beta\epsilon_i}
\left<\psi_i\middle|\hat{\sigma}\middle|\psi_i\right>}
{\sum_ie^{-\beta\epsilon_i}}\\
=&
\epsilon_0+\frac{e^{-\beta\left(\epsilon_1-\epsilon_0\right)}}{g}\sum_{i=g}^{g+g_1-1}
\left<\psi_i\middle|\epsilon_1\hat{\sigma}-\epsilon_0\middle|\psi_i\right>
+\\
&o\left(e^{-\beta\left(\epsilon_1-\epsilon_0\right)}\right).\raisetag{0pt}
\end{split}
\end{equation}
Putting Eqs.~\eqref{zqz1N1} and \eqref{dzratio} into Eq.~\eqref{eq:pfpT}, we can get the conclusion that all the coefficients in Eq.~\eqref{eq:pfpT} are 
of order $\beta^2e^{-\beta\left(\epsilon_1-\epsilon_0\right)}$ or smaller.
Therefore, $\left.\partial f/\partial T\right|_{\xi=z_i}\to0$ and $dz_i/dT\to0$.
The corresponding term associated with $dz_i/dT$ will vanish at 0~K if $z_i\mp1$ is nonzero.
For bosonic systems, all the $z_i-1$ are nonzero.
Thus, we have
\begin{equation}
	\begin{split}
		\label{dz1dTb}
		-&\lim_{T\to0}\left[
		k_B\ln Z(\beta,\xi=+1)+k_BT\frac{d\ln Z_{N^1}}{dT}\right]\\
		=&\lim_{T\to0}\left[\frac{F(\beta,\xi=+1)}{T}-k_BT\frac{d\ln Z_{N^1}}{dT}\right]=0.
	\end{split}
\end{equation}
However, for fermionic systems, $z_1+1$ equals 0.
The derivative of the free energy with respect to the temperature satisfies
\begin{equation}
\begin{split}
\label{dz1dTf}
&-k_B\ln Z(\beta,\xi=-1)-k_BT\frac{d\ln Z_{N^1}}{dT}-k_BT\frac{dz_1/dT}{z_1+1}\\
&=\frac{F(\beta,\xi=-1)}{T}-k_BT\frac{d\ln Z_{N^1}}{dT}-k_BT\frac{dz_1/dT}{z_1+1}\to0.
\raisetag{-12pt}
\end{split}
\end{equation}
Now, if we substitute Eq.~\eqref{dz1dTb} into \eqref{dz1dTf}, we can get
\begin{equation}
\label{dz1dT}
k_BT\frac{dz_1/dT}{z_1+1}=\frac{E_{0}+o(1)}{T},
\end{equation}
where $E_0\equiv F_{0f}-F_{0b}$ is the difference between the 0~K free energies of bosonic/fermionic systems with the same potential energies and the difference only comes from exchange symmetry.
Here, $F_{0b/f}\equiv F(T=0~{\text{K}},\xi=\pm1)=-\lim_{T\to0}k_BT\ln Z(\beta,\pm1)$.
Then, by solving Eq.~\eqref{dz1dT}, we have
\begin{equation}
\label{z1}
z_1=-1+e^{-\beta E_0+o(\beta)},
\end{equation}
for the description of the evolution of the zero $z_1$ when the temperature deviates from the 0~K limit.
The existence of the last term in Eq.~\eqref{dFdT} for the fermionic system means that the behaviors of the fermionic system's free energy 
is different from those of the bosonic one and other anyons from the perspective of the mathematical expressions of them. 
Specifically, for the fermionic case, because $z_1=-1$, $\left.\dot{z_1}/(-1-z_1)\right|_{\beta\to\infty}$in Eq.~\eqref{dFdT} will not vanish and it will 
contribute a finite energy $E_0$. 
Therefore, the zero-point energy of the fermionic system has a different analytic form from that of the bosons, even if the potentials are the same.
This contrast lies at the core of thermodynamics.
It indicates that there is a transition between the bosonic and fermionic systems, which has similar behaviors of a ``phase transition'' at 0~K, due to 
the fact that the analytic features of the thermodynamic laws governing the bosonic world and fermionic world are different.
This can rationalize the failures of the existing studies of the FSP based on analytic continuations at low temperatures.
In subsequent papers, we will provide results on the evolution of the LY zeros of $\xi$ at finite temperatures and seek recipe to circumvent the FSP.

\section{Conclusion remarks}
\label{Conclusion}
In conclusion, we establish an analytic form of the partition function of the indistinguishable particle systems with a complex $\xi$.
This polynomial of the partition function as a function of $\xi$ has a simple form at this 0~K limit.
With this form, we locate these LY zeros.
The one at $\xi=-1$ is special since it corresponds to the fermonic system.
This LY zero of $\xi$ induce an extra term for the free energy of the fermonic system in comparison with that in the bosonic system and quantum systems composed by other anyons.
The different analytic form of the free energy means that there is a transition similar to phase transition between the quantum world composed by fermions and bosons, even 
if their potentials are the same.
With these, we provide a fresh perspective to understand the successes/failures of existing studies of FSP based on analytic continuations.
\newline

\begin{acknowledgments}
    %
    We are supported by the National Science Foundation of China under Grant Nos. 12234001, 12204015, 12474215, 62321004, and the National Basic Research Programs of China under 
    Grand Nos. 2021YFA1400500 and 2022YFA1403500.
\end{acknowledgments}
\bibliography{ref}

\begin{thebibliography}{64}
\bibitem [{cod()}]{cod_and_fsp}%
  \BibitemOpen
  \href@noop {} {}\bibinfo {note} {The FSP and the curse of dimensionality are
  closely related. In PIMD/PIMC, the curse of dimensionality is only
  circumvented for distinguishable particles. For realistic indisginguishable
  particles, this is not yet the case.}\BibitemShut {Stop}%
\end{thebibliography}
\end{document}